\documentclass[aip,graphicx,preprint]{revtex4-1}

\draft 
\usepackage{graphicx}
\usepackage{dcolumn}
\usepackage{bm}

\usepackage[cmex10]{amsmath}
\usepackage{amssymb}
\usepackage{gensymb}
\usepackage{subfigure}

\begin{document}





\title{Analysis of volume distribution of power loss in ferrite cores} 








\author{M. LoBue}
\author{V. Loyau}
\author{F. Mazaleyrat}





\affiliation{SATIE, ENS de Cachan, CNRS, UniverSud, 61 av du President Wilson, F-94230 Cachan, France}





\date{\today}

\begin{abstract}

We present a technique to estimate the inhomogeneities of magnetic loss across the section of ferrite cores under AC excitation. The technique is based on two distinct calorimetric methods that we presented elsewhere. Both the method are based on the measurement of the rate of increase of the sample temperature under adiabatic condition. The temperature ramp is recorded either measuring the sample bulk resistivity, or using a platinum probe pasted on the sample surface. As an example we apply the procedure to an industrial sample of Mn-Zn ferrite under controlled sinusoidal excitation with a peak induction of $50$ mT in the range between $100$kHz and $2$MHz. The results are discussed by comparison with simulations of the dissipation field profile through the sample, calculated using a FEM code.

\end{abstract}

\pacs{75.50.-y, 75.50.Gg, }

\maketitle 



\section{Introduction} \label{SEC-I}

Soft sintered Mn-Zn ferrites are widely used in high frequency applications based on switching circuits. Their relatively high resistivity and permeability make them optimal for the $30$kHz-$1$MHz bandwidth. 
Miniaturization and the increase of working frequencies make thermal and loss managements the keys for future design of electronic switching devices. It is thus necessary to establish reliable methods to measure, understand and predict the losses in soft ferrites at medium and high frequencies. 

Losses are commonly measured using the flux-metric method \cite{Sato1987-1, Saotome1997-1, Fiorillo2009-1}. However, when excitation frequencies are above $10$ kHz, measurements  can be dramatically affected by spurious phase contributions due to stray capacitances and inductances of the circuits and probes. A possible alternative is the calorimetric method; in adiabatic conditions, the power dissipated in the magnetic material is directly related to the heating rate, reducing the measurement to that of a temperature ramp. We discussed in detail this technique in a recent paper  \cite{Loyau2009-1} where an experimental set-up  using a platinum probe as temperature sensor was presented. A further improvement of the technique has been proposed \cite{Loyau2010-1}   by measuring the temperature from the bulk DC resistivity of the ferrite core \cite{Loyau2010-1}. Whereas the temperature deduced from bulk resistivity represents a measurement averaged over the whole sample volume (as in the case of flux-metric measurements), the platinum probe measurement gives information on dissipation taking place in a thin strip of the material just near its surface. Consequently the comparison between the two measurements can be used  to estimate the loss inhomogeneity, and thus, the relevance of skin effect for a given sample geometry and excitation frequency. It will be shown that, global low level electromagnetic  parameters, obtained by impedance spectroscopy, can be used as local input parameters for the FEM computation of magnetic loss in linear regime. In this way we shall calculate the volume or surface averaged loss values, corresponding to the two types of calorimetric measurements and compare them with experimental results. 

\section{Experimental and computation details} \label{SEC-II}

Power losses in magnetic materials under AC excitation are associated with all the irreversible phenomena taking place during the core magnetisation (i.e. Barkhausen jumps, eddy currents, domain wall resonance, spin dumping, magnetoacoustic emission, etc.). Under  adiabatic conditions \cite{Loyau2009-1}, all the dissipated power is converted into heat and induces a temperature increase of the sample. Thus the temperature field through the core $T(x, y, z, t)$ can be described by means of the linearised heat equation giving:

\begin{equation} \label{EQ:dT/dt}
\frac{\partial T(x, y, z, t)}{\partial t} = \frac{1}{\rho_m c_p} p_s(x, y, z)
\end{equation}

where $p_s(x, y, z)$ is the local heat source density (in Wm$^{-3}$), $c_p$ is the specific heat (in JK$^{-1}$kg$^{-1}$) of the sample and $\rho_m$ its density (in kgm$^{-3}$). Now, if $V_0$ is the volume of the sample, the magnetic loss per cycle measured using the flux-metric method, at a given excitation frequency $f$, corresponds to the average loss over the sample volume, namely:

\begin{equation} \label{EQ:W}
 \langle W \rangle = \frac{c_p}{f V_0} \int_V \frac{\partial T(x, y, z, t)}{\partial t} dV.
\end{equation}

Whereas a measure of the average temperature of the whole sample will result in a correct extrapolation of $\langle W \rangle$, any local measurement will be related with dissipations taking place just in a small volume near the position of the temperature probe. In the latter case  $\langle W \rangle$ can be deduced properly only when volume losses are uniformly distributed through the sample. 

Two different techniques where used to measure the temperature ramp: the first  using a platinum probe pasted (Pt500) on the sample \cite{Loyau2009-1} and therefore sensible to the temperature change in a narrow strip just near the surface, and a second technique  based on the measurement of previously calibrated bulk resistivity of the ferrite core \cite{Loyau2010-1} which returns a global average of the temperature ramp. The latter leads $\langle W \rangle$ notwithstanding the inhomogeneities of the loss field due to the geometry of the sample and to the skin effect. In both cases, the sample was a square cross section bar (I rod: $0.64\times0.64\times2.54$ cm$^3$) made of 3E27 Ferroxcube Mn-Zn ferrite. Complex impedance spectra of the permeability $\mu = \mu' + i \mu''$ and of the electrical conductivity $\sigma = \sigma' + i \sigma'' = \sigma' + i \epsilon \omega$ (where $\epsilon$ is the electrical permittivity and $\omega =  2 \pi f$ is the angular frequency) where performed with a HP 4195 impedance analyser between 100 kHz and 10 MHz. The losses where measured in a vacuum chamber between 100 kHz and 2 MHz as described in ref. \cite{Loyau2009-1,Loyau2010-1} from $10$ second long temperature recordings.



In order to validate the interpretation of the measurements performed with the two techniques, an approach based on the solution of Maxwell equations was used. For this purpose, the code FEMM\cite{FEMM} was used taking  measurements of complex magnetic permeability and conductivity as intrinsic material parameters. All the following analysis is based on three strong assumptions: (\textit{i}) the peak induction values are sufficiently low to consider the magnetic behaviour as linear. Thus the concept of real and imaginary magnetic permeability and their relationship with energy losses is valid. This assumption is confirmed by the fact that all the loss measurements presented in this paper scale roughly as the square of the peak induction. This means that the term $\mu''/\mid \mu \mid^2$ remains constant in the investigated induction range (from $3$mT to $50$mT);
(\textit{ii}) all the measurements of complex magnetic permeability are performed on a toroidal ferrite sample with section small enough to make eddy current loss and skin effect negligible; 
(\textit{iii}) all the measurements of complex conductivity are performed on bar shaped samples which section is sufficiently reduced to neglect skin effect. 

From (\textit{ii}) the measured $\mu$ is a parameter related to the local phase lag between $\textbf{H}$ and $\textbf{B}$ due to all dissipative phenomena but eddy current losses. On the other hand, the complex electrical conductivity $\sigma$ results from the mixed electrical properties of the grain and of the grain boundaries \cite{Goodenough2002-1}. Its real part $\sigma'$, is related to ohmic and dielectric loss effect whereas the imaginary part $\sigma''$, is due to the effect of mixed permittivity.

\begin{figure}
\centering
\subfigure[]{\includegraphics[scale=0.5]{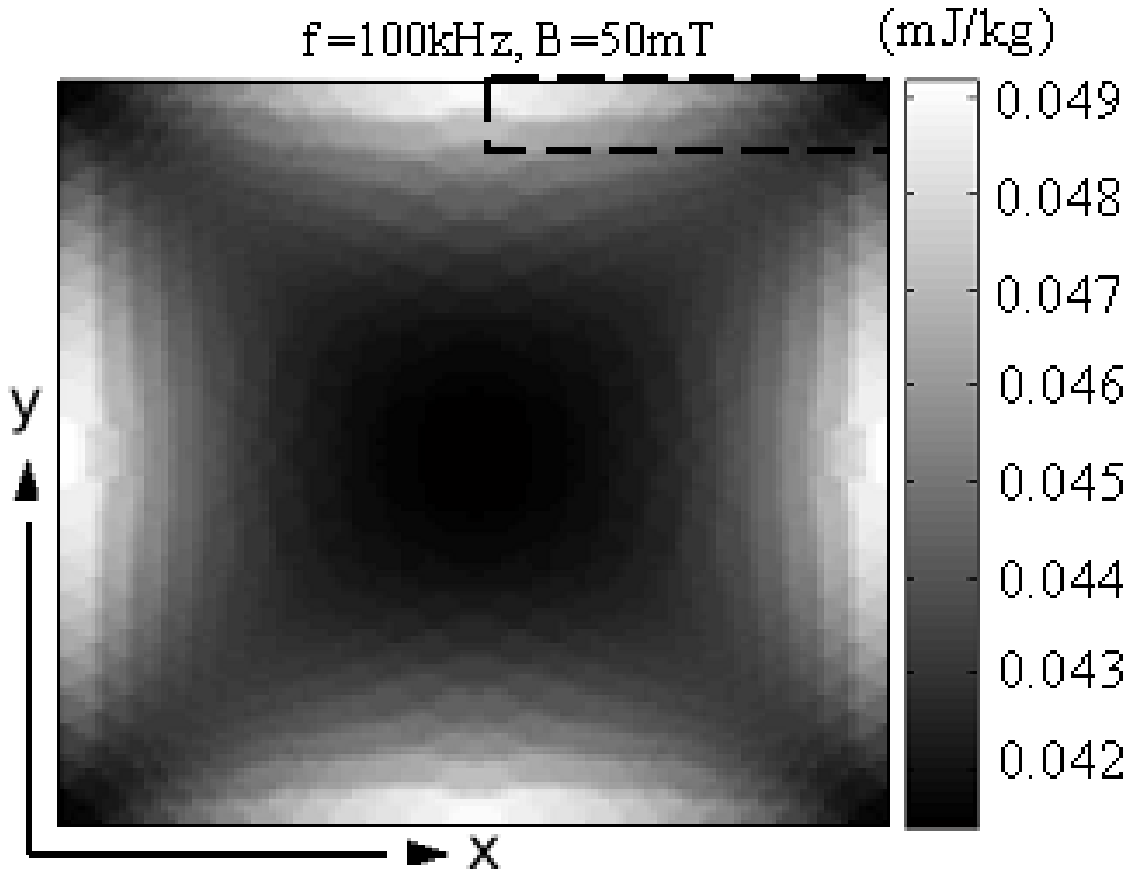}
\label{fig1a}
} 
\centering
\subfigure[]{\includegraphics[scale=0.5]{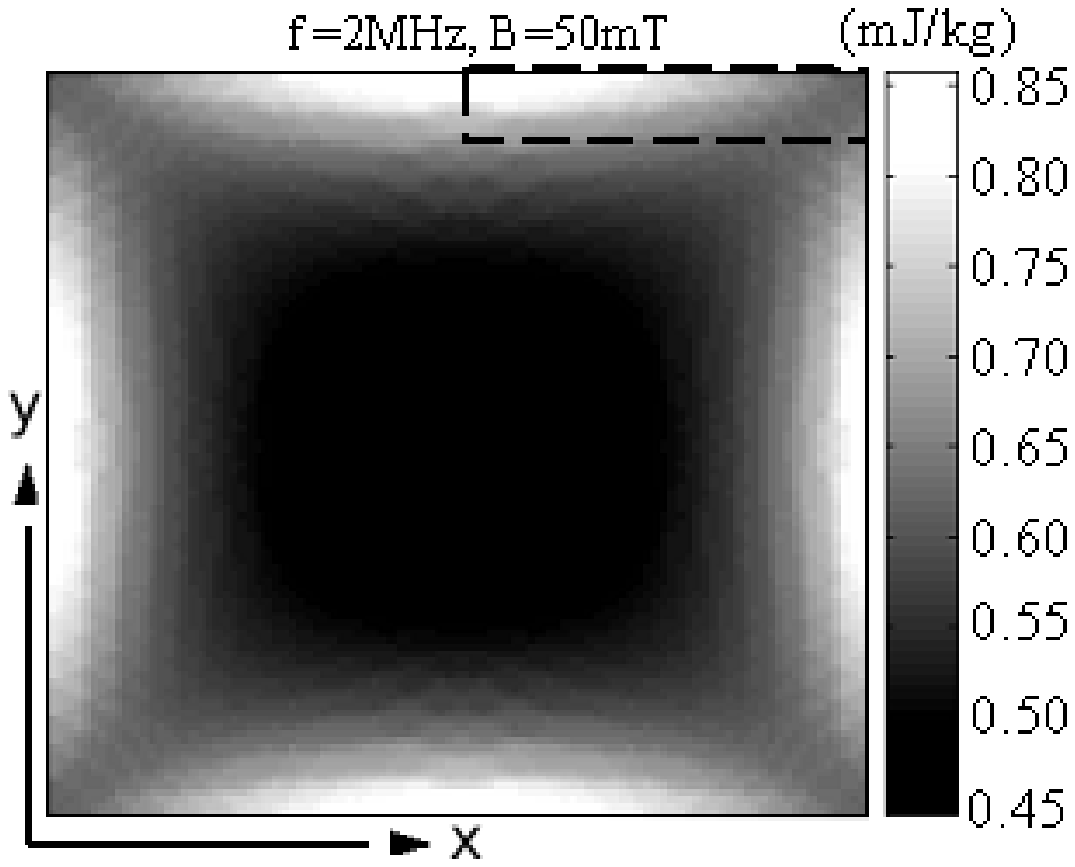}
\label{fig1b}
}
\caption[Optional caption for list of figures]{Total loss field $w_{tot}(x, y)$ calculated from Eq. (\ref{EQ:EDDY}) and (\ref{EQ:magloss}) for $f=100$kHz \subref{fig1a} and f=$2$MHz \subref{fig1b}. The dashed rectangle indicates the region used to calculate the surface loss as measured by the platinum probe. The thickness of the rectangle is defined as the product of the heat propagation speed in the material and the measure duration ($10$ seconds). Grey scales indicate the local loss in mJkg$^{-1}$ where the loss varies from $0.04$mJkg$^{-1}$ (dark regions) to  $0.05$mJkg$^{-1}$ (white regions) in \subref{fig1a} , and  from $0.45$mJkg$^{-1}$  to $0.85$mJkg$^{-1}$  in \subref{fig1b} . The ratio between the maximum  and the minimum values is $1.25$ at $f=100$kHz \subref{fig1a} and $2$ at f=$2$MHz \subref{fig1b}}
\label{fig1}
\end{figure}

\begin{figure}
\includegraphics[scale=0.65]{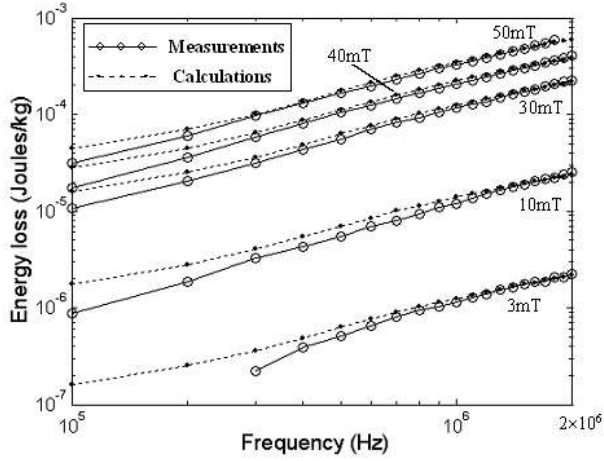}
\caption{Comparison between the bulk measurements of the average total loss and the calculated ones. Peak inductions are: $3$mT, $10$mT, $30$mT, $40$mT and $50$mT  }
\label{fig2}
\end{figure}

\begin{figure}
\includegraphics[scale=0.65]{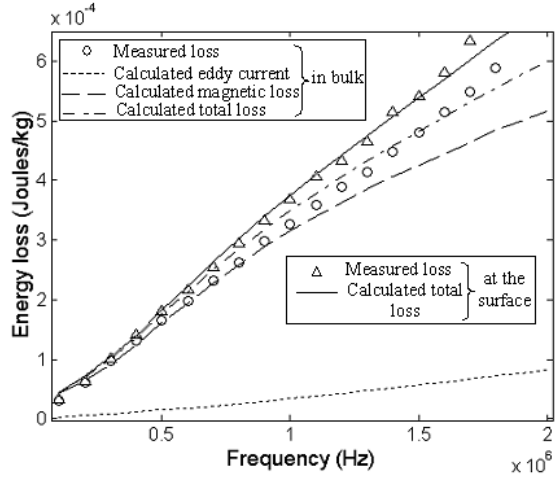}
\caption{Comparison between bulk and surface loss measurements and the same for computations (over the whole volume and the one delimited by the dashed line in Fig.\ref{fig1}.  The peak induction is of $50$mT  }
\label{fig3}
\end{figure}

Given the geometry of the problem, with the magnetic field always perpendicular (along $z$ axis) to the core section, the calculation of the loss field reduces to a two dimensional problem in the $xy$ plane. Now, under sinusoidal excitation when $H_z(x, y, t) = H_0(x, y) e^{i \omega t}$, substituting the measured $\sigma$ and $\mu$ in the diffusion equation for the magnetic field we have,

\begin{equation} \label{EQ:Diffusion2}
  \frac{1}{\mu} \triangledown^2 H_z(x, y, t) = i \omega \sigma H_0(x, y) .
\end{equation}

From this equation and from the fact that $ \textbf{j} = \triangledown \times \textbf{H}$ we can calculate the eddy-current loss field across the core section,

\begin{equation}\label{EQ:EDDY}
 p_{ec}(x, y) = \frac{1}{2} \rho' \lvert \textbf{j}(x, y) \rvert^2
\end{equation}

where $\rho' = Re(1/\sigma)$. This expression includes all the eddy current related losses. The effect of ohmic and dielectric dissipation in both, the grains and their boundaries, are described by the real part of the resistivity $\rho'$.
On the other hand, the complex part of the permeability $\mu''$, measured on the very small sample, will account for all the magnetisation process related dissipation phenomena ( excess losses, spin damping, etc.). So we can write the magnetic loss field as:

\begin{equation}\label{EQ:magloss}
 p_{mag}(x, y) = \frac{1}{2} \omega \mu'' \lvert \textbf{H}_z (x, y) \rvert^2= \frac{1}{2} \omega \frac{\mu''}{\lvert \mu \rvert^2 } \lvert \textbf{B}_z(x, y) \rvert^2
\end{equation}

The total power loss will be the sum of the two terms of Eq. (\ref{EQ:EDDY}) and (\ref{EQ:magloss}), and the loss per cycle and unit mass will be $ w_{tot}(x, y)= p_{tot}(x, y) / (f \rho_m)$. By averaging over the core section and multiplying by the length, we obtain $\left\langle W\right\rangle_{v}=\ell\int_{-a/2}^{a/2}\int_{-a/2}^{a/2}w(x,y)\mathrm{d}x\mathrm{d}y$ corresponding to the bulk (resistance) measurement, whereas $\left\langle W\right\rangle _{s}=\ell\int_{0}^{a/2}\int_{0}^{d}w(x,y)\mathrm{d}x\mathrm{d}y$ yields the value corresponding to surface (PT 500) measurement. The thickness $d$ is defined as the product  of the heat diffusion velocity in the material and the duration of the measurement ($10$ seconds). 

\section{Results and discussion} \label{SECC-IV}

In order to determine the magnetic field $H_z(x, y)$ across the section area we must specify the induction conditions under which losses have been both calculated and measured. Here we consider the case where a sinusoidal magnetic induction of average value $\langle B_0 \rangle$ is forced across the core section area $S$. This implies that the magnetic field verifies:

\begin{equation}\label{EQ:Induction}
 \langle B_0 \rangle = \frac{1}{S} \int_S \mu H_0(x, y) dS
\end{equation}

The magnetic field $H_z(x, y)$ has been determined solving Eq.(\ref{EQ:Diffusion2}) coupled with Eq.(\ref{EQ:Induction}) with a two dimensional finite element code in a frequency range between $100$kHz and $2$MHz on the sample geometry. Calculation results are shown in Fig. \ref{fig1} where the loss field $w_{tot}(x, y)$ is plotted for $f=100$kHz in Fig.\ref{fig1a} and $f=2$MHz in Fig.\ref{fig1b}.

In Fig.\ref{fig2} comparison between the bulk measurements of loss and the calculated average total loss $\left\langle W\right\rangle_{v}$ is shown for peak inductions ranging from $3$mT to $50$mT. The agreement between measurements and calculations is particularly good for frequencies above $300$kHz. The over estimation of computed losses may be due to the compensation error in the measurement of $\mu''$ performed with the impedance analyser. Indeed, at low frequency the resistive part of the impedance is nearly zero, so even a small absolute error can yield a strong systematic error. 
Fig.\ref{fig3}  shows the comparison between the surface measured loss and corresponding computation $\left\langle W\right\rangle_{s}$. Calculations and measurements are in good agreement. With the sample geometry studied here, as the average magnetic path is not much different from the extremal paths in the sample, all the loss inhomogeneities can be ascribed to the skin effect. So, either measurements and finite elements calculations, show that skin effect is negligible in a frequency regime below $600$-$700$kHz. Above these frequencies skin effect becomes relevant, and a large part of the dissipation phenomena is concentrated near the surface of the sample. 
In conclusion, it has been shown that the calorimetric loss measurement method, exploiting bulk and surface temperature recordings, is able to give indirect information on the eddy current losses in Mn-Zn ferrites. Using FEM computation --for which material parameters input where obtained by impedance spectroscopy-- it is possible to compute the loss fields and to obtain very good agreement with measurements without introducing any fitting parameter. Hence, this model could be integrated in magnetic components design software in order take into account the magnetic losses including eddy currents for the dimensioning.









%




%











%













\begin{thebibliography}{1}
\bibitem{Sato1987-1}
T. Sato, and Y. Sasaki, IEEE Trans. Magn. \textbf{23}, 2593 (1987)
\bibitem{Saotome1997-1}
H. Saotome, and Y. Sakaki, IEEE Trans. Magn. \textbf{33}, 728 (1997)
\bibitem{Fiorillo2009-1}
F. Fiorillo, M. Co\"{\i}sson, C. Beatrice, and M. Pasquale, J. Appl. Phys. \textbf{105}, 07A517 (2009) 
\bibitem{Loyau2009-1}
V. Loyau, M. LoBue, F. Mazaleyrat, Rev. Sci. Instrum. \textbf{80}, 024703 (2009)
\bibitem{Loyau2010-1}
V. Loyau, M. LoBue, F. Mazaleyrat, IEEE Trans. Magn. \textbf{46}, 529 (2010)
\bibitem{FEMM}
Finite Element Method Magnetics, D. Meeker, freeware, http://www.femm.info
\bibitem{Goodenough2002-1}
J. B. Goodenough, IEEE Trans. Magn. \textbf{38}, 3398 (2002)

\end{thebibliography}
\end{document}